\documentclass[twocolumn,pre,aps,showpacs]{revtex4}
\newcommand{\BEQ}{\begin{equation}}     
\newcommand{\BEA}{\begin{eqnarray}}
\newcommand{\EEQ}{\end{equation}}       
\newcommand{\EEA}{\end{eqnarray}}
\def\be{\begin{equation}}
\def\ee{\end{equation}}
\def\bc{\begin{center}}
\def\ec{\end{center}}
\newcommand{\D}{{\rm d}}

\begin{document}

\input epsf.sty




\input epsf.sty

\widetext

\title{Reply to "Comment on 'Scaling of the linear response in simple 
ageing systems without disorder' "}

\author{Malte Henkel$^1$ and Michel Pleimling$^2$}
\affiliation{$^1$Laboratoire de Physique des Mat\'eriaux (CNRS UMR 7556),
Universit\'e Henri Poincar\'e Nancy I, B.P. 239,
F -- 54506 Vand{\oe}uvre l\`es Nancy Cedex, France\\
$^2$Institut f\"ur Theoretische Physik I,
Universit\"at Erlangen-N\"urnberg, D -- 91058 Erlangen, Germany}

\begin{abstract}
The value of the non-equilibrium exponent $a$ is measured in the
two-dimensional ($2D$) Ising model quenched to below 
criticality from the dynamical scaling of 
the zero-field-cooled and the intermediate susceptibility. Our results fully 
reconfirm the expected value $a=1/2$ but are inconsistent with the value
$a=1/4$, advocated by Corberi, Lippiello and Zannetti 
({\tt cond-mat/0506139}).

\end{abstract}
\pacs{05.20.-y, 64.60.Ht, 75.40.Gb}
\maketitle
For simple magnets quenched to below their critical point, one generally 
expects, in the scaling regime where $t,s\gg t_{\rm micro}$ and
also $t-s\gg t_{\rm micro}$ ($t_{\rm micro}$ is a microscopic reference time),
the following scaling behaviour of the linear autoresponse function
\BEQ \label{R}
R(t,s) = \left. \frac{\delta \langle\phi(t)\rangle}{\delta h(s)}\right|_{h=0} 
\sim s^{-1-a} f_R(t/s)
\EEQ
where $\phi(t)$ is the order parameter at the observation time $t$ and $h(s)$
is the 
magnetic field at the waiting time $s$. 

For some time,
Corberi, Lippiello and Zannetti (CLZ) \cite{Corb05,Corb01} have 
advocated a phenomenological formula
\BEQ \label{Corberi}
a = \frac{n}{z} \left( \frac{d-d_L}{d_U-d_L}\right)
\EEQ
where $d_{U,L}$ are the upper and lower critical dimensions, respectively and
$n$ is the number of components of the order parameter. 
On the other hand, the commonly accepted physical picture \cite{Cate00} of the 
ageing (coarsening) process going on after the quench has it that
ordered domains form very rapidly and that the ageing comes from the movement
of the domain walls between the ordered regimes. If this idea is combined with
scaling arguments and if one takes into account that in certain systems (such
as the Ising model and referred to as class S) the spatial correlations decay 
exponentially while in others (such as the spherical model and referred to as 
class L) the spatial correlations decay algebraically, viz. 
$C_{\rm eq}(\vec{r}\,)\sim |\vec{r}\,|^{-(d-2+\eta)}$, 
one obtains \cite{Henk03,Henk04}
\BEQ \label{gl:a}
a = \left\{ \begin{array}{ll}
                          1/z          & \mbox{\rm ~; for class S} \\
                          (d-2+\eta)/z & \mbox{\rm ~; for class L}
\end{array} \right. 
\EEQ
where $z$ is the dynamical exponent. 
The result (\ref{gl:a}) has been reproduced in many studies. However, if
eq.~(\ref{Corberi}) of CLZ should turn out to be correct 
that would also invalidate the
simple physical picture of the coarsening process mentioned above. 
A good practical way to decide between (\ref{Corberi}) and the prediction (\ref{gl:a}) appears to be a study of the $2D$ Ising model with a non-conserved 
order parameter and quenched to $T<T_c$ from a fully disordered initial state.
Here $z=2$ is known \cite{Bray94} and from eqs.~(\ref{Corberi}) and (\ref{gl:a}) one has $a=1/4$ and $a=1/2$, respectively. 

CLZ first arrived at eq.~(\ref{Corberi}) by analyzing 
numerical data \cite{Corb01} of the
field-cooled susceptibility 
$\chi_{\rm ZFC}(t,s) = \int_{s}^{t}\!\D u\, R(t,u)$. From a straightforward
integration of (\ref{R}), they obtained $\chi_{\rm ZFC}(t,s) \sim s^{-a}$
and proceeded to extract $a$. However, it was pointed out that 
$\chi_{\rm ZFC}(t,s)$ does not obey a simple scaling but rather there may be
a further and dominant contribution coming from the upper limit of integration.
For systems of class S, there are well-defined domains with domain walls whose
thickness is small with respect to the domain size and one has \cite{Henk04} 
\BEQ \label{chizfc}
\chi_{\rm ZFC}(t,s) = \chi_0  + \chi_1 t^{-A} - s^{-a} f_M(t/s) \;\; ; \;\;
\mbox{\rm class S}
\EEQ
where $\chi_{0,1}$ are constants and $f_M(x)\geq 0$ is a scaling function. 
Indeed, it can be shown that $A=z^{-1}-\kappa$ \cite{Henk04}, where the 
exponent $\kappa\geq 0$ describes the time-dependent scaling of the width of 
the domain walls $w(t)\sim t^{\kappa}$ \cite{Abra89}. In the $2D$ Ising model
$\kappa=1/4$ is known \cite{Abra89}, hence $A=1/4$ which explains the early
result of CLZ but it becomes clear that the term na\"{\i}vely expected
from the scaling behaviour (\ref{R}) merely yields a finite-time correction. 
CLZ did not take into account the condition $t-s\gg t_{\rm micro}$ for the
validity of the scaling form (\ref{R}) in their analysis \cite{Corb01} and 
hence have missed the leading time-dependent term in (\ref{chizfc}). 

On the other hand, for systems of class L, we argued previously that because
of the long-range correlations, the effective width of the domains should be
the same as their linear size which leads to $A=0$ \cite{Henk04}. Since the
clusters have no `inside', the term analogous to $\chi_0=0$ in 
eq.~(\ref{chizfc}) is absent, but it's r\^ole is taken
over by the constant $\chi_1$. CLZ \cite{Corb05} point out correctly that 
for both classes S and L the constant 
$\lim_{t\to\infty}\chi_{\rm ZFC}(t,s)$ (with $x=t/s$ fixed) and which
equals $\chi_0$ for class S and $\chi_0+\chi_1$ for class L can
be related to the equilibrium magnetization through the static 
fluctuation-dissipation theorem. In summary one has, with $a$ given 
by (\ref{gl:a})
\BEQ \label{final}
\hspace{-2truemm}\chi_{\rm ZFC}(t,s) = \left\{ \begin{array}{ll}
\frac{1}{T}
\left( 1 - m_{\rm eq}^2\right) 
+ \chi_1 t^{-A} - s^{-a} f_M(t/s) & \\
 & \hspace{-2.5truecm} \mbox{\rm ~; for class S} \\
\frac{1}{T}\left( 1 - m_{\rm eq}^2\right) - s^{-a} f_M(t/s) & \\
 & \hspace{-2.5truecm} \mbox{\rm ~; for class L}
\end{array} \right. 
\EEQ

\begin{figure}
\centerline{\epsfxsize=3.1in\epsfbox
{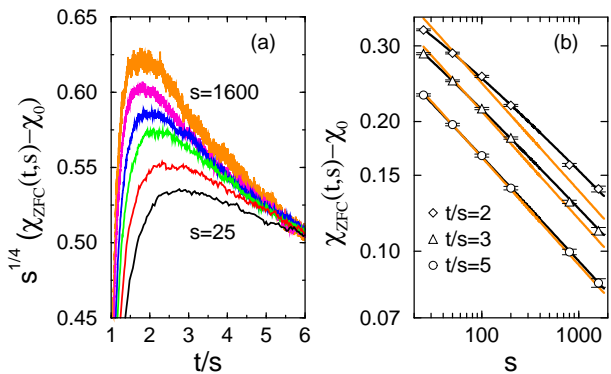}
}
\caption{(a) Scaling plot of the reduced field-cooled susceptibility 
$\chi_{\rm ZFC}(t,s)-\chi_0$ against $t/s$ in the $2D$ Ising model 
quenched to $T=1.5$, for waiting times $s=25$, 50, 100, 200, 800, 1600
(from bottom to top). 
(b) Comparison of the data with a fit $\chi_{\rm ZFC}(t,s)-\chi_0=
a_0 s^{-1/4} + a_1 s^{-1/2}$ (black lines) and a fit
$\chi_{\rm ZFC}(t,s)-\chi_0= a_0 s^{-1/4}$ (grey lines), 
for several values of $t/s$. 
For clarity the data for $t/s=2$ and for $t/s=3$ 
have been shifted upwards by a factor 1.4 and 1.2, respectively.
\label{Abb1}} 
\end{figure}

In their comment, CLZ \cite{Corb05} now argue that the exponents
$A$ and $a$ were really not distinct and that one rather had $a=A$. From now
on, we concentrate on systems of class S and furthermore on the $2D$ Ising model
quenched below $T_c$. 
{}From numerical simulations, CLZ reproduce once more that
$\chi_{\rm ZFC}(t,s)-\chi_0\sim s^{-1/4}$, as generally expected. But they do
not consider explicitly the corrections to that leading scaling behaviour and
merely claim, but do not prove, the absence of the scaling corrections
of order $s^{-a}$, which are expected from eq.~(\ref{final}). 

Indeed, the scaling behaviour of $\chi_{\rm ZFC}$ is not as simple as claimed 
by CLZ. We show this in figure~\ref{Abb1}a, with data obtained from the
standard heat-bath method. It can be seen that 
the collapse in this scaling plot is far from complete which strongly indicates
the presence of sizeable finite-time corrections. The origin of these
becomes clear in figure~\ref{Abb1}b where we 
compare our data to the asymptotic form $\chi_{\rm ZFC}(t,s)-\chi_0 = 
a_0 s^{-1/4} + a_1 s^{-1/2}$, for several values of $t/s$. While taking into
account both terms does reproduce the data well, we included for comparison
also the fits where $a_1=0$ was fixed by hand, as suggested by CLZ. It is clear
that the data, which display a pronounced curvature,
can only be fitted if $a_1<0$, as expected from (\ref{final}). 
Indeed, the best fits yield $a_0=0.68$ and $a_1=-0.35$ for $t/s=2$,
$a_0=0.62$ and $a_1=-0.19$ for $t/s=3$ and 
$a_0=0.54$ and $a_1=-0.06$ for $t/s=5$ in excellent quantitative
agreement with (\ref{final}). 
We conclude that there is no indication in favour of an absence of the 
last term in eq.~(\ref{final}), hence $a>A$ which provides clear 
counterevidence against the proposal of CLZ. 

We had already
given earlier an analysis of the scaling of the thermoremanent
magnetization (where terms of order $s^{-A}$ do not occur). By
taking into account also the leading finite-time correction we found
full compatibility with $a=1/2$ but inconsistency with
$a=1/4$ \cite{Henk03}. 

\begin{figure}[!h]
\centerline{\epsfxsize=3.1in\epsfbox
{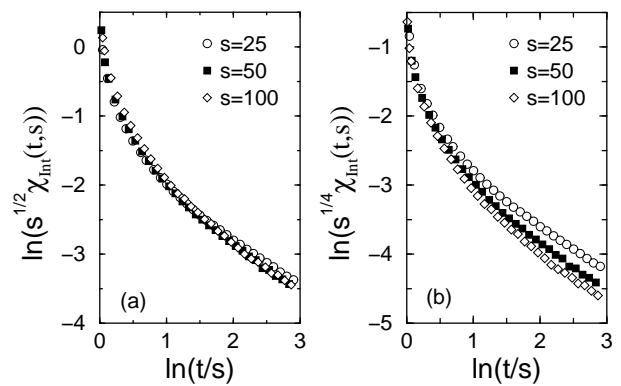}
}
\caption{Scaling plot of the intermediate susceptibility $\chi_{\rm Int}(t,s)$
for (a) the assumed value $a=1/2$ (b) the assumed value $a=1/4$.
Error bars are smaller than the symbol sizes.
\label{Abb2}} 
\end{figure}

We may also show that indeed $a=1/2$ in the $2D$ Ising model through the
so-called intermediate integrated response \cite{Henk04,Pleim05}
\BEQ
\chi_{\rm Int}(t,s) = \int_{s/2}^{s} \!\D u\, R(t,u) 
\sim s^{-a} f_{\rm Int}(t/s)
\EEQ
which has the advantage that the leading term $\sim t^{-A}$ which dominates
over the scaling term in $\chi_{\rm ZFC}$ is absent. In figure~\ref{Abb2} we 
examine the scaling of $\chi_{\rm Int}(t,s)$ and try to achieve a collapse for
$a=1/2$ and $a=1/4$. While there is a nice collapse for $a=1/2$ already for
relatively small values of the waiting time (unless $t/s$
is too close to unity), the data fail to collapse for $a=1/4$. Again, this is
fully consistent with $a=1/2$ but excludes $a=1/4$. 

Turning to systems of class L, CLZ nicely reconfirmed the expected scaling
forms eqs.~(\ref{gl:a},\ref{final}). However, there is nothing in their test
with contradicts the simple scaling we used in \cite{Henk04} to obtain $A=0$.  

In conclusion, having reexamined the scaling of some integrated susceptibilites,  
we have shown that the scaling of $\chi_{\rm ZFC}(t,s)$ does
indeed contain at least two important contributions, 
see eq.~(\ref{final}), which is
against the proposal of CLZ. We stress
that a simple demonstration of scaling of $\chi_{\rm ZFC}$ is not enough to be
able to reliably know which of the exponents $A$ or $a$ is measured. It is
more safe to study a quantity such as the intermediate susceptibility, which
does not suffer from this difficulty. Applied to the Ising model quenched to
$T<T_c$, our results fully confirm eq.~(\ref{gl:a}) but disagree 
with eq.~(\ref{Corberi}). 

This work was supported by CINES Montpellier (projet pmn2095).
MP was supported by the Deutsche Forschungsgemeinschaft through 
grant no. PL 323/2.


\end{document}